\documentclass[11pt]{article}
\RequirePackage[OT1]{fontenc}
\usepackage{amsthm,amsmath,amssymb,amsfonts,bm,enumerate,xcolor,graphicx,natbib,color,url}
\RequirePackage[colorlinks,citecolor=blue,urlcolor=blue]{hyperref}
\usepackage{hyperref}
\usepackage{subcaption} 
\usepackage{ulem}
\usepackage{pifont}
\usepackage{textcmds}
\usepackage{booktabs}
\usepackage{multirow}
\usepackage{booktabs}
\usepackage{adjustbox}
\usepackage[usestackEOL]{stackengine}
\usepackage[toc,page]{appendix} 
\usepackage{algorithm,algorithmic}
\usepackage{lipsum}
\usepackage{setspace}

\usepackage{setspace}
\allowdisplaybreaks
\def\frac#1#2{{\textstyle{#1\over#2}}}

\DeclareSymbolFont{AMSb}{U}{msb}{m}{n}
\DeclareMathSymbol{\Natural}{\mathbin}{AMSb}{"4E}
\DeclareMathSymbol{\Integer}{\mathbin}{AMSb}{"5A}
\DeclareMathSymbol{\Real}{\mathbin}{AMSb}{"52}
\DeclareMathSymbol{\Rational}{\mathbin}{AMSb}{"51}
\DeclareMathSymbol{\Imaginary}{\mathbin}{AMSb}{"49}
\DeclareMathSymbol{\Complex}{\mathbin}{AMSb}{"43} 
\DeclareMathSymbol{\Disk}{\mathbin}{AMSb}{"44} 
\def\bi{\begin{itemize}}
\def\ei{\end{itemize}}
\def\bd{\begin{description}}
\def\ed{\end{description}}
\def\ben{\begin{enumerate}}
\def\een{\end{enumerate}}
\def\bv{\begin{verbatim}}
\def\ev{\end{verbatim}}

\def\2to{{\ {\buildrel 2\over \longrightarrow}\ }}

\def\I1ton{{$I_1,\ldots,I_n$}}
\def\X1ton{{$X_1,\ldots,X_n$}}
\def\Y1ton{{$Y_1,\ldots,Y_n$}}
\def\Z1ton{{$Z_1,\ldots,Z_n$}}
\def\R1ton{{$R_1,\ldots,R_n$}}
\def\e1ton{{$e_1,\ldots,e_n$}}
\def\t1ton{{$t_1,\ldots,t_n$}}
\def\x1ton{{$x_1,\ldots,x_n$}}
\def\y1ton{{$y_1,\ldots,y_n$}}
\def\z1ton{{$z_1,\ldots,z_n$}}
\newcommand{\blind}{1}

\begin{document}
%%%%%%%%%%%%%%%%%%%%%%%%%%%%%%%%%%%%%%%%%%%%%%%%%%%%%%%%%%%%%%%%%%%%%%%%%%%%%%
\thispagestyle{empty}
\baselineskip=28pt
\vskip 5mm

\renewcommand{\thefootnote}{\fnsymbol{footnote}}

\begin{center} 
{\Large{\bf Statistics of extremes for natural hazards: landslides and earthquakes}}\footnotemark[2]
\end{center}

\baselineskip=12pt

\vskip 5mm

\renewcommand{\thefootnote}{\arabic{footnote}}

\if1\blind
{
\begin{center}
\large
Rishikesh Yadav$^1$, Luigi Lombardo$^2$, and Rapha\"el Huser$^3$
\end{center}
\renewcommand{\thefootnote}{\fnsymbol{footnote}} \footnotetext[2]{Extract of Chapter 27 (first 5 pages), to be published in the \emph{Handbook of Statistics of Extremes}, eds.\ Miguel de Carvalho, Rapha\"el Huser, Philippe Naveau and Brian Reich}
\renewcommand{\thefootnote}{\arabic{footnote}} \footnotetext[1]{
\baselineskip=10pt  Department of Decision Sciences, HEC Montr\'eal, Montr\'eal, Canada. E-mail: rishikesh.yadav@hec.ca}
\footnotetext[2]{
\baselineskip=10pt  ITC, Faculty of Geo-Information Science and Earth Observation, University of Twente, Enschede, The Netherlands. E-mail: l.lombardo@utwente.nl}
\footnotetext[3]{
\baselineskip=10pt Statistics Program, Computer, Electrical and Mathematical Sciences and Engineering (CEMSE) Division, King Abdullah University of Science and Technology (KAUST), Thuwal 23955-6900, Saudi Arabia. E-mail: raphael.huser@kaust.edu.sa}
\fi

\baselineskip=26pt
\vskip 2mm
\centerline{\today}
\vskip 4mm

%%%%%%%%%%%%%%%%%%%%%%%%%%%%%%%%%%%%%%%%%%%%%%%%%%%%%%%%%%%%%%%%%%%%%%%%

\baselineskip=14pt

In this chapter, we illustrate the use of split bulk--tail models and subasymptotic models motivated by extreme-value theory in the context of hazard assessment for earthquake-induced landslides. A spatial joint areal model is presented for modeling both landslides counts and landslide sizes, paying particular attention to extreme landslides, which are the most devastating ones.

\section{Introduction}
%\subsection{Background and motivation}
Statistics of extremes has been used extensively in climate science for the modeling of low-probability high-impact events, including natural hazards such as heavy precipitation \citep{katz2002statistics,cooley2007bayesian,huser2014space,deFondevilleDavison18}, extreme heatwaves \citep{Davison.Gholamrezaee:2012,winter2016heatwaves,Zhong.etal:2021,thompson2023most,Zhang2023explaining}, and strong windstorms or hurricanes \citep{opitz2016modeling,Risser2017Attributable,Dawkins2018,huser2021maxid}, among others; such extreme-value models are often developed with the ultimate ambition to improve the state-of-the-art in performing hazard and/or risk assessment and mitigation, and in attributing specific catastrophic extreme events to human influence under climate change. By contrast, the application of extreme-value theory (EVT) and statistics to the modeling and prediction of geophysical processes, such as devastating landslides or earthquakes, is sparser in the literature, perhaps due to the different type of data involved and the added modeling difficulties that come with it. Unlike climate data that are often measured at fixed locations (e.g., monitoring stations, or on a spatial grid with climate model outputs) and regular intervals (e.g., hourly or daily) over a period of time, the exact location and timing of landslides and earthquakes are typically unknown before they occur---they are thus often treated as random; moreover, the exact conditions and triggering mechanisms often differ for each event, which implies that the data are rarely replicated (unless some kind of spatio-temporal aggregation is used), making any extreme-value analysis more challenging, and that they typically require other kinds of specialized statistical models such as point processes \citep{moller1998log,illian2008}. Nevertheless, statistical methods based on EVT have still been used in data-driven geophysical sciences, e.g., for estimating the maximum earthquake magnitude \citep{Beirlant2019Estimating,darzi2023Bayesian}, for probabilistic seismic hazard analysis \citep{dutfoy2022randomized}, as well as for landslide hazard mapping over space \citep{yadav2023joint,dahal2024junction}; see also \cite{Kiriliouk2019peaks} who fit a multivariate extreme-value model to extreme rainfall data, in order to (indirectly) understand the probability of rainfall-induced landslides through the rainfall intensity--duration thresholds for landslide initiation established by \cite{guzzetti2007rainfall}. 

In this chapter, we focus on the spatial modeling of earthquake-induced landslides \citep{lombardo2019geostatistical}, and present a general Bayesian hierarchical modeling framework that has been used in the literature (in various forms and contexts, and with some variations) for jointly modeling multiple occurrences and sizes of various natural phenomena over space \citep{pimont2021prediction,koh2021spatiotemporal,yadav2023joint}. While all (big or small) future landslides matter for appropriate hazard assessment, 
\emph{extreme} landslides are particularly devastating and should therefore be given strong attention in modeling. In the model construction pipeline, we thus advocate using probability distributions motivated by EVT that can adequately capture the upper tail behavior, while simultaneously providing a good fit in the bulk. Specifically, we here review and compare two distinct approaches for modeling the full range of landslide sizes: (i) split (also called `mixture', `spliced', `piecewise', or `hybrid') bulk--tail models \citep{Behrens.etal:2004,Carreau.Bengio:2009,Scarrott.MacDonald:2012,opitz2018inla,castro2019spliced,pimont2021prediction,koh2021spatiotemporal}, wherein the upper tail is modeled with the asymptotically-justified \emph{generalized Pareto (GP)} distribution (see \cite{Davison.Smith:1990,Davison.Huser:2015} and Chapter~2 for more details) and the bulk is modeled using another, essentially arbitrary distribution; and (ii) `subasymptotic' distributions (see, e.g., \cite{Naveau.etal:2016,yadav2021spatial,yadav2021flexible,Stein:2021a,Stein:2021b}, and Chapter~5) capturing the lower tail and the upper tail flexibly in compliance with EVT with a smooth transition in between. While there are several possible subasymptotic distributions, we focus in this chapter on the \emph{extended generalized Pareto (eGP)} distribution \citep{Papastathopoulos.Tawn:2013}, which is a parsimonious parametric model that has found various applications in the study of natural hazards including extreme precipitation \citep{Naveau.etal:2016}, wildfires \citep{cisneros2024deep}, and landslides themselves \citep{yadav2023joint}. To streamline the analysis and the discussion, we here consider only one of the simplest eGP distributions, but see Chapter~5 for more details on other possible options and extensions. From a hazard assessment perspective, both landslide sizes and landslide counts (modeled using the Poisson distribution) are important (see Section~\ref{sec:hazard} for details), and they are here modeled jointly using a latent Gaussian model, in which fixed effects and shared spatially-structured effects can be easily incorporated at the latent level to capture complex spatial trends, spatial and cross-dependencies, as well as covariate effects. Joint modeling of occurrences and sizes is especially key for properly assessing the uncertainty of hazard estimates, which are obtained as a function of both elements.

In this chapter, we showcase the versatility of the proposed general modeling framework using an inventory comprising thousands of landslides simultaneously initiated by the devastating 2008 Wenchuan earthquake in China, going beyond \cite{lombardo2019geostatistical} who studied the spatial distribution of landslides triggered by the same earthquake but did not consider the modeling of landslide sizes. Incorporating physical knowledge into probability models is important to improve both model fit and interpretability, and is one of the recommendations listed in the opinion piece by \cite{huser2024modeling}; using informative geophysical and geomorphological covariates, and defining the model at the `slope unit' resolution (more details in Section~3), are two possible ways to achieve this goal. Therefore, in our case study, both landslides and covariates are observed at the slope unit level, thus requiring a joint areal model for occurrences and sizes, instead of a continuous-space marked point process model. This chapter aims at illustrating this extreme-value-based modeling framework for the study of landslide data, and to compare the pros and cons of split bulk--tail models with subasymptotic models in practice. While we find no uniformly-better approach in this case, we argue that the eGP distribution has an overall better performance and is a more natural modeling solution in general. We also discuss how to perform scalable Bayesian inference in this framework based on a customized efficient Markov chain Monte Carlo (MCMC) algorithm, coupled with latent random effects with a sparse precision matrix (here, chosen with an intrinsic conditional autoregressive (iCAR) probabilistic structure; see \cite{besag1974spatial}). The selection of highly informative covariates (e.g., peak ground acceleration) that can potentially replace the use of latent spatial effects is also discussed, and various latent model structures are compared.
 
The rest of this chapter is structured as follows: Section~\ref{sec:hazard} gives a more precise definition of the `hazard' according to the geoscience literature. Section~3 provides details on the Wenchuan landslide data inventory used in our case study. Section~4 presents the modeling framework and MCMC-based Bayesian inference. Section~5 discusses results for the Wenchuan data application. Finally, Section~6 concludes with a summary of some key points.

\section{Hazard definition}\label{sec:hazard}
Although processes such as floods, earthquakes, or landslides largely differ in their physical manifestation, geophysical and statistical properties, and their impacts, international guidelines on natural hazard prediction share some core requirements \citep{mitchell1993natural}. Broadly speaking, hazard prediction should reflect \emph{where} and \emph{when} (or how frequently) a future natural hazard might occur, and if it does, \emph{how big} (or how destructive) it might be. We also stress that in the geoscience literature, `risk assessment' has a specific meaning that goes beyond `hazard assessment' by additionally encapsulating the \emph{exposure} and \emph{vulnerability} of infrastructure and/or people at risk, and estimating the associated economic, societal and/or environmental costs. In this chapter, we thus distinguish these different notions and adopt the terminology commonly used in geoscience.

The first requirement (`where') involves the notion of \emph{susceptibility}, which indicates how likely a given portion of the landscape is to undergo a given hazard \citep{karlsson2017natural,nicu2022multi}. Statistical models used to address this question in the literature include logistic regression models for presence-absence data, Poisson regression models for count data, or more advanced point process models such as log-Gaussian Cox processes (see, e.g., \cite{lombardo2018point,lombardo2020space}). 

The second requirement (`when' or `how frequently') relates to the \emph{return period} of a given hazard and reflects its recurrence over time. The frequency of such devastating phenomena is often difficult to model because temporal replicates are rarely available and rich spatio-temporal inventories are scarce. Moreover, landslides often occur as part of a sequential compound extreme event (e.g., after a major earthquake or heavy rain), which also complicates the estimation of their return period as this requires an understanding of the frequency of the main triggering factor(s). This aspect of hazard assessment is, therefore, often the most neglected one (but see \cite{dahal2024junction} for a recent attempt to take it into account in a study of rainfall-induced landslides). 

The third requirement (`how big' or `how destructive') involves the notion of \emph{intensity} of an event, which indicates the energy and level of threat associated with a given hazard when it occurs \citep{peng2005space,hungr2018some}. 
Note this notion of intensity differs from that classically used in statistics with point process models. 

In the context of landslides, David J.~Varnes and the International Association of Engineering Geology, Commission on Landslides and Other Mass Movements on Slopes defined the landslide hazard more specifically through statistical terms as \textit{``the probability of occurrence within a specified period and a given area of a potentially damaging phenomenon''} \citep{varnes1984}. This definition was later updated by \cite{guzzetti1999landslidehazard} to explicitly include the intensity of the event. While this general definition was proposed in the context of landslides, the same formulation can be interpreted more broadly and applied similarly in other contexts with various types of natural hazards.

In Section~4, we present an extreme-value-based latent Gaussian modeling pipeline that complies with the first and third requirements of the hazard definition. That is, the proposed model can be used to jointly estimate both the susceptibility and the intensity of a natural phenomenon---here, in the context of earthquake-induced landslides. As there are different ways to characterize the intensity of a natural phenomenon, there is no clear consensus in the literature on how to represent and model it. For instance, flood intensity \citep{vojtek2016flood} is usually expressed as a function of the water height (e.g., \cite{van2023breakthrough}) or peak flow (the maximum rate of discharge; \cite{formetta2021assessment}). 
Earthquake intensity \citep{hough2014earthquake} is commonly expressed as the peak of either the displacement \citep{trugman2019peak}, velocity \citep{dahal2023ground}, or acceleration \citep{murphy1977correlation} experienced at a given location. 
Similarly, landslide intensity \citep{lari2014probabilistic} can also be expressed in multiple ways: for instance, the velocity \citep{he2023modelling} or kinetic energy \citep{pudasaini2021mechanics} associated with a failing mass, the force that the landslide body may exert onto an object \citep{tang2014novel}, its volume \citep{jaboyedoff2020review}, planimetric area \citep{DiNapoli2023} or other shape indices \citep{rana2023landslide}, are all accepted as valid ways to describe the landslide intensity. 
However, most of these elements cannot be measured in the context of large landslide populations. 
Velocities over large landscapes are measurable through Interferometric Synthetic Aperture Radar (InSAR) but only for slow movements \citep{Ahmad2023}. 
Kinetic energy and force can only be estimated through demanding numerical simulations \citep{pudasaini2022landslide}, and their real measurement is conditional on the availability of very expensive instruments installed on a slope for monitoring purposes \citep{mazzanti2015new}. 
Similarly, landslide volumes cannot be measured unless topographic data are available before and after the failure \citep{tseng2013application}. 
Among these landslide intensity parameters, the landslide area (or derivatives thereof) is the easiest one to collect because it is a byproduct of any rigorous landslide mapping procedure \citep{lombardo2021landslide}.    
Hence, out of all the metrics listed above, the landslide area is the most relevant one to support  statistical analyses. In the case study presented in this chapter, we model the square root of the landslide area (easily measurable through remote sensing), which roughly describes the `diameter' of the affected area and can be understood as a proxy for the landslide size and its destructiveness.

\baselineskip 12pt
\bibliographystyle{CUP}
\bibliography{ref}

\end{document}